\documentclass[a4paper,11pt]{article}
\usepackage{pos}

\title{Parallel tempering algorithm applied
to the deconfinement transition of
quenched QCD }
\ShortTitle{Parallel tempering algorithm applied
to quenched QCD }

\author*[a]{Ruben Kara}
\author[a]{Szabolcs Borsányi}
\author[a,b,c,d]{Zoltán Fodor}
\author[b]{Daniel A. Godzieba}
\author[a,b]{Paolo Parotto}
\author[e]{Dénes Sexty}

\affiliation[a]{University of Wuppertal, Department of Physics, Wuppertal D-42119, Germany}
\affiliation[b]{Pennsylvania State University, Department of Physics, State College, PA 16801, USA}

\affiliation[c]{Jülich Supercomputing Centre, Forschungszentrum Jülich, Jülich D-542425, Germany}
\affiliation[d]{E{\"o}tv{\"o}s University, Budapest 1117, Hungary}
\affiliation[e]{University of Graz, Department of Physics, Graz A-8010, Austria}

\emailAdd{rkara@uni-wuppertal.de}

\abstract{
QCD with infinite heavy quark masses exhibits a first-order thermal transition which is driven by the spontaneous breaking of the global $\mathcal{Z}_3$ center symmetry. We analyze the corresponding order parameter, namely the Polyakov loop and its moments, and show, with a rigorous finite size scaling, that in the continuum limit the transition is of first order. We show that the use of a parallel tempering algorithm can significantly reduce the large auto-correlation times which are mainly caused by the supercritical slowing down. As a result, we calculate the transition temperature $w_0 T_c$ with per-mill precision, and the latent heat, carrying out controlled continuum and infinite volume extrapolations.
}

\FullConference{%
The 39th International Symposium on Lattice Field Theory,\\
8th-13th August, 2022,\\
Rheinische Friedrich-Wilhelms-Universität Bonn, Bonn, Germany
}

\begin{document}
\maketitle

\section{Introduction}	
\noindent
In the case of physical quark masses, it is well known that the QCD thermal transition for vanishing chemical potential is an analytic crossover \cite{Aoki:2006we}. For infinite heavy quark masses, QCD is reduced to a pure gauge theory in which the quarks are static and can only propagate in time. In this scenario QCD exhibits a 1st order thermal transition due to the spontaneous breaking of the underlying global $\mathcal{Z}_3$ center symmetry. This symmetry breaking is related to the (de-)confinement transition since the  the corresponding order parameter, the Polyakov loop $P$, is linked to the free energy of a static quark. In the confined phase, an infinite amount of energy is needed to remove a quark from the system and $| \langle P \rangle |=0.$  In contrast to that, in the deconfined phase, $| \langle P \rangle |$ takes a finite non-vanishing value. One main feature of a 1st order phase transition is the existence of a latent heat which was calculated recently in the continuum limit for quenched QCD \cite{Shirogane:2016zbf, Shirogane2021}. Here the key to calculate the latent heat is the separation of the configurations in hot and cold "phases". In these regimes the trace anomaly is evaluated and the difference is taken, yielding the latent heat up to a normalization factor. We extend this work by calculating the latent heat in the continuum and infinite volume limit, and by applying a parallel tempering algorithm in $\beta$, originally applied to spin models \cite{Swendsen1986,Marinari1992}. Several extensions followed with applications to lattice QCD \cite{joo1998,ilgen2002,boyd1998}, which treated the quark masses, coupling and the hopping parameter (or their combinations) as dynamical variables. Progress could also be achieved for the problem of topological freezing  and twist-sectors \cite{burgio2007,hasenbusch2017,bonati2021, borsanyi2021}. In our case, parallel tempering in $\beta$ allows us to perform a high precision study of the deconfinement transition of quenched QCD since it lowers tremendously the high auto-correlation times caused by supercritical slowing down. Simulating real phase transitions on finite systems gets more and more severe as the volume is increased since the system tends to stick in one phase. By collecting contributions from other ensembles at different couplings this effect can be significantly reduced. We calculate the transition temperature $w_0 T_c$ with per-mill precision, the trace anomaly and latent heat. This contribution is mainly based on our work \cite{prec_study}.

\section{Analysis}
\noindent
In quenched QCD the Polyakov loop $P$ works as a true order parameter to probe the center symmetry breaking. $P$ transforms non-trivially under $\mathcal{Z}_3$ and can be defined as
\begin{equation}
P=\frac{1}{N_s^3} \sum_{\vec{x}} P_{\vec{x}} = \frac{1}{N_s^3} \sum_{\vec{x}} \mathrm{tr} \left[ \prod_\tau U_4(\vec{x},\tau)   \right], 
\end{equation}
where $N_s$ stands for the spatial extension of the lattice and $\vec{x}$ and $\tau$ indicate the spatial and temporal position respectively.\\
In the case of a 1st order phase transition $P$ should show a discontinuity at the critical coupling $\beta_c$ in the thermodynamic limit, whereby its susceptibility $\chi$ diverges linearly with the physical volume. For a 2nd order phase transition this behavior is accompanied by a critical exponent. Hence, analyzing the peak of the susceptibility determines the type of phase transition and the corresponding transition temperature $T_c$. Another way to extract $T_c$ is the zero-crossing of the third-order Binder cumulant $b_3$ of the absolute value of the Polyakov loop. The susceptibility and the Binder cumulant are defined as
\begin{equation}
 \chi=N_s^3 \left( \langle |P|^2 \rangle   - \langle |P| \rangle^2 \right),  \qquad	b_3=\frac{  \langle  (|P| - \langle |P| \rangle)^3 \rangle   }{  \langle  (|P| - \langle   |P|\rangle^2  \rangle^{3/2}    }.
\end{equation}
\noindent
The third Binder cumulant is a measure of the skewness of the $|P|$ distribution. Simulating far away from $\beta_c$, i.e. where the system is deep in one phase, there is a single (Gaussian-) peak in the distribution. As one comes closer and closer to the critical coupling, a second peak evolves which corresponds to the 2nd phase of the system. As soon as $\beta_c$ is reached, both peaks have the same shape and the distribution is completely symmetric which leads to a  vanishing skewness. While strictly speaking there are no phases in a finite system, we loosely use the word phase to describe the (de-)confined regimes.


\section{Parallel tempering to improve on supercritical slowing down}
\noindent
Simulating real phase transitions makes us face the phenomenon of critical, in the case of 2nd order transitions, and supercritical slowing down for 1st order phase transitions. Both cause a high auto-correlation times, which one can significantly reduce by employing parallel tempering in $\beta$ \cite{Marinari1992}. The mechanisms behind critical and supercritical slowing down are similar, but not completely identical. In the case of a 1st order phase transition various quantities show a discontinuity at $T=T_c$. For SU(3) quenched QCD the effective potential of the Polyakov loop contains three degenerate deconfined minima and one confined minimum. These statements are only true in the thermodynamic limit. So in a finite volume a temperature range around $T_c$ exists, in which the system can be in both phases and tunnel between them. These states are split by the energy $\Delta E$, which is proportional to the volume $V$. Since both phases have to be sampled, this gets more severe as $V$ is increased. 
\begin{center}
\includegraphics[width=0.65\textwidth]{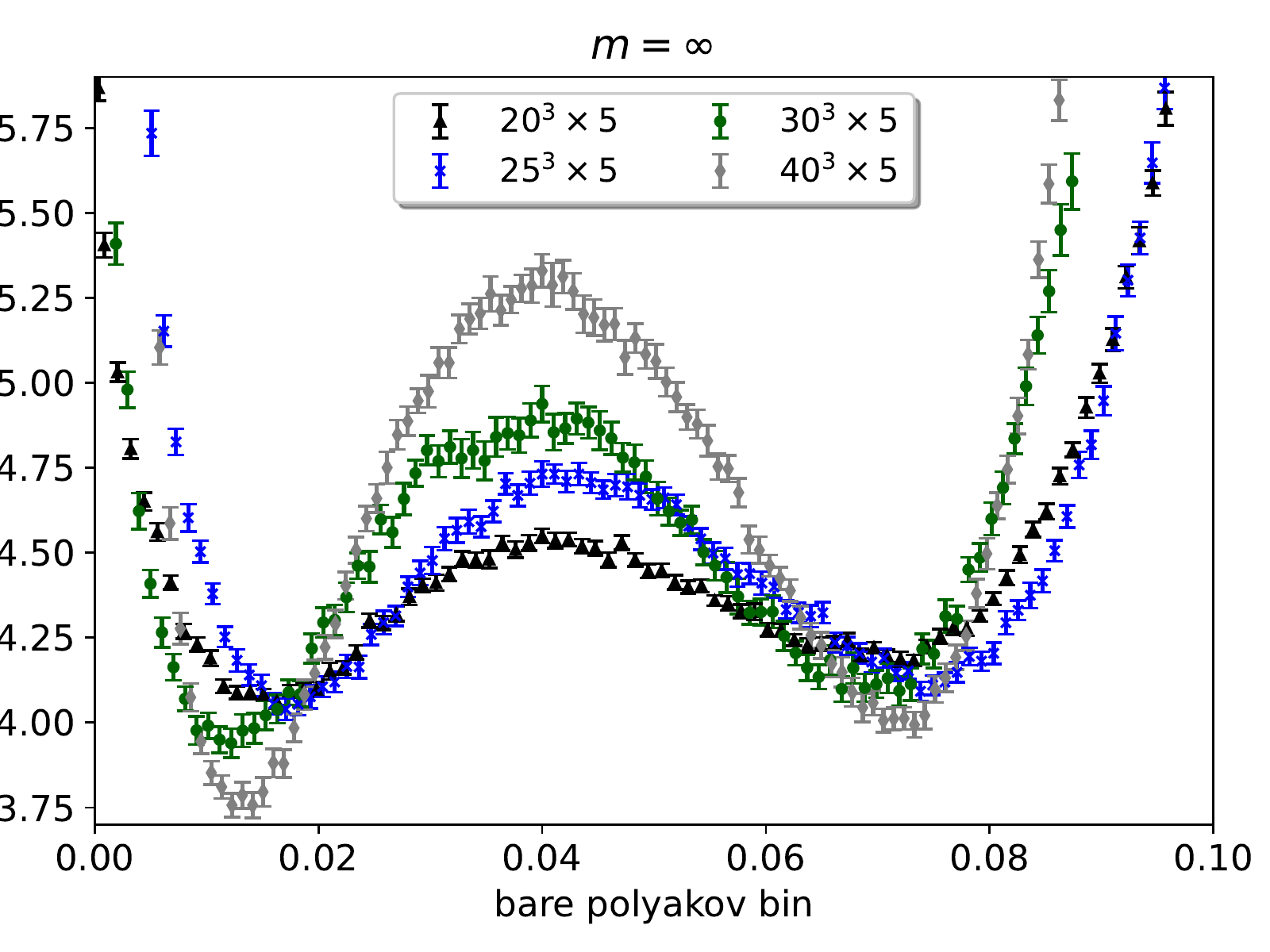}
\captionof{figure}{Effective potential of the absolute value of the Polyakov loop for various volumes in quenched QCD, after employing parallel tempering. The barrier between both phases increases with the volume.}
\label{fig:potential_quenched}
\end{center}
\noindent
The consequence of this is a high auto-correlation time if tunneling is not sampled, since the system tends to stick in one phase. In Fig. \ref{fig:potential_quenched} the effective potential of the absolute value of the Polyakov loop is shown and one can see that the barrier between the two phases increases with the volume.\\
In the case of a 2nd order phase transition a diverging correlation length is expected, which leads to a high auto-correlation time as well. In both cases, it is possible to reduce the auto-correlation time tremendously by using the parallel tempering algorithm, thus performing multiple simulations at different $\beta$. These are distinct Markov processes with an overlapping equilibrium distribution. Parallel tempering adds to the "standard" Markov transitions within these single sub-ensembles another Metropolis accept/reject step to swap configurations between pairs of them, with a swapping probability $P_s$ given by 
\begin{align}
&P_s(i,j)=\mathrm{min}\left(1,e^{-\Delta \mathcal{H} } \right),\\
&\Delta \mathcal{H}=\left(\mathcal{H}_j(a) + \mathcal{H}_i(b) \right) - (\mathcal{H}_i(a) + \mathcal{H}_j(b) ).
\end{align}
\noindent
Here $i$, $j$ indicate two sub-ensembles and $a$, $b$ their configurations, respectively \cite{joo1998}. This means that swapping configurations is more likely for neighboring ensembles. As a result the auto-correlation time within the sub-ensembles is reduced, since they gather contributions of the other ensembles at different couplings. Hence both phases of the system around $\beta_c$ can be sampled. The price to pay is a resulting correlation between the sub-ensembles. In Fig. \ref{fig:pt_vs_bf} the advantages of parallel tempering compared to standard brute force simulations are clearly visible. With the same amount of computer time, the statistical errors of the tempering results are smaller and describe much better a high statistic result, especially close to $\beta_c$. 
\begin{center}
\includegraphics[width=0.65\textwidth]{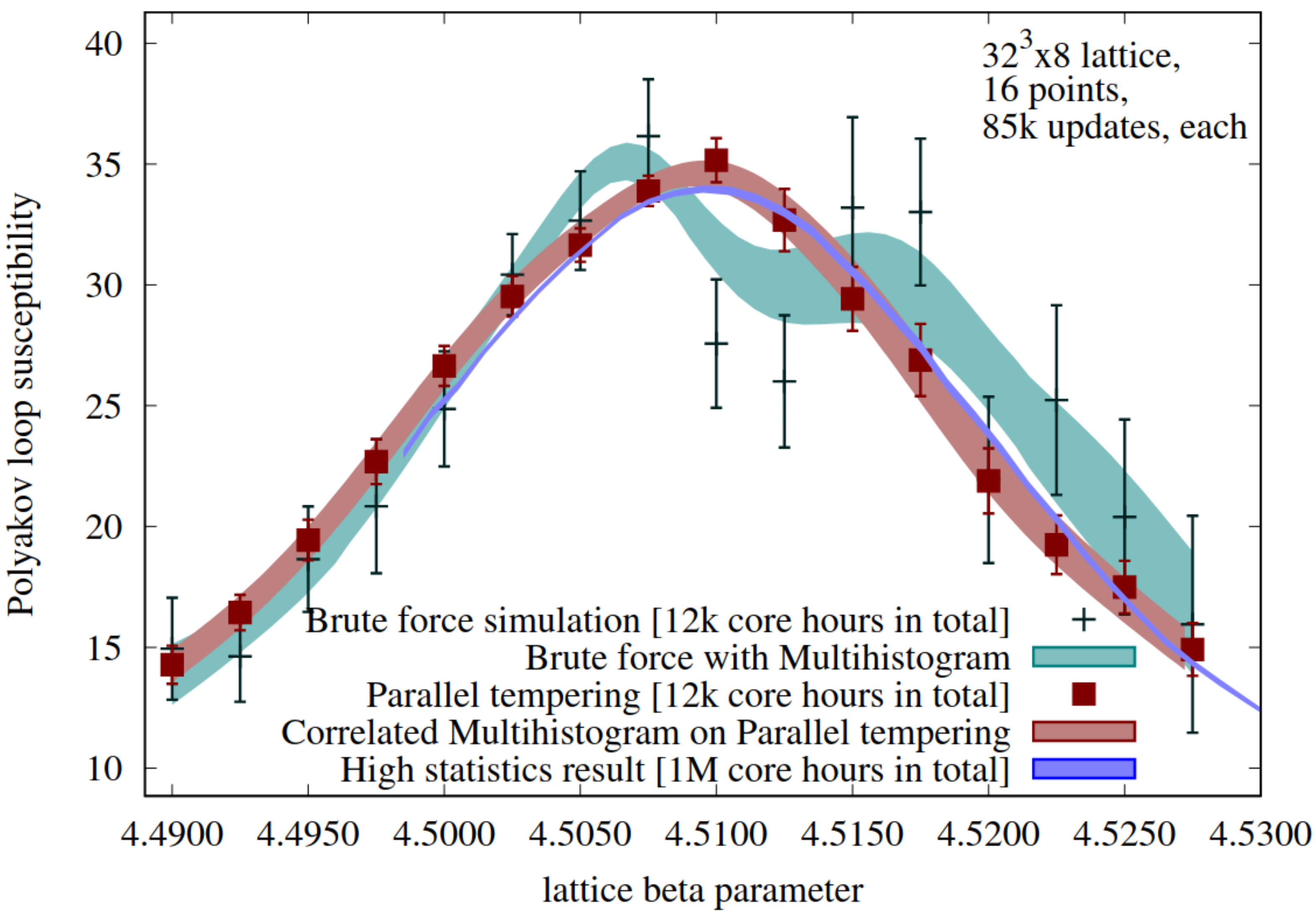}
\captionof{figure}{Polyakov loop susceptibility as a function of the coupling $\beta$ in quenched QCD on a $32^3 \times 8$ lattice. The red squares correspond to parallel tempering simulations and the red band indicates the interpolation between them obtained by the correlated multihistogram method. Compared to the standard brute force simulations (black crosses) with the same amount of computer time, they give a much better description of the high statistic result (blue band).}
\label{fig:pt_vs_bf}
\end{center}
\noindent	
For parallel tempering simulations the distance between neighboring ensembles $\Delta \beta
$ and their total number $n$ play a crucial role. Increasing $\Delta \beta$ suppresses the probability to swap configurations and the ensembles collect more contributions from their neighbors until they "decouple" and run as independent simulations. For a suitable acceptance rate of swapping updates, the action distributions of neighboring ensembles should clearly overlap, which can be obtained by tuning $n$ and $\Delta \beta$. In Fig. \ref{fig:auto_corr_times} the auto-correlation time of the Polyakov loop for different sets of parallel tempering simulations is compared to standard methods. The lowest auto-correlation times are achieved with the densest and largest $\beta-$grid.
\begin{center}
\includegraphics[width=0.75\textwidth]{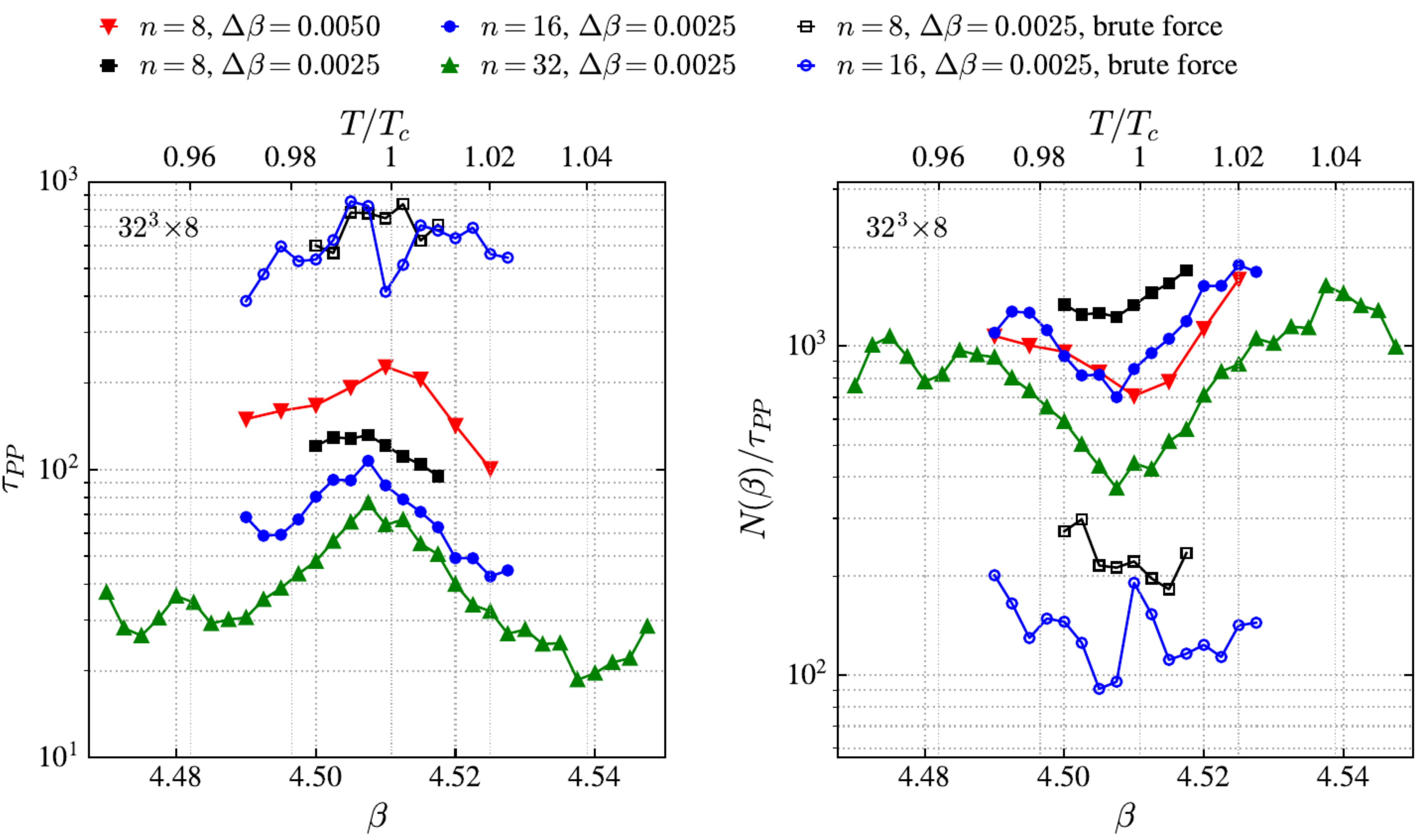}
\captionof{figure}{Left panel: Auto-correlation time of the Polyakov loop as a function of the coupling $\beta$ on a $32^3 \times 8$ lattice using parallel tempering (filled points) and brute force simulations (unfilled points) with the same amount of computer time. $\Delta \beta$ indicates the $\beta$-spacing between the sub-ensembles and $n$ stands for the total number of them. Right panel: Number of statistically independent configurations divided by the auto-correlation time as a function of $\beta$.}
\label{fig:auto_corr_times}
\end{center}
\noindent
It is important to note, that the major driver to the auto-correlation time at one $\beta$ value comes from instances in which the same stream contributes. Therefore we order the Monte Carlo chain according to the stream ID number, whereby the chronological order of each stream is kept. This way the contributions of each single stream are brought together additionally minimizing the correlation between blocks in the jackknife analysis.

\section{Transition temperature and latent heat}
\noindent
To relate the action dependent value of $\beta$ to a more generic transition temperature, it is necessary to set the scale. In this study we use $w_0$  based on the Wilson flow \citep{borsanyi2012, luescher2010, Fodor:2014cpa}. Measured in lattice units $w_0/a$, we find no significant volume dependence, since the effects are comparable with the statistical errors. With this quantity a transition temperature can be defined as
\begin{equation}
w_0  T_c = \frac{w_0}{a N_t}(\beta_c).
\end{equation} 
\noindent
In the infinite volume limit, the definition of $\beta_c$ by the peak of the susceptibility or vanishing $b_3$ should not play any role. Both results have to agree in the thermodynamic limit, since the peak of $\chi$ diverges linearly, the width vanishes linearly and the slope of $b_3$ (close to $\beta_c$) increases linearly with the volume. 
This behavior is clearly shown for the transition temperature in Fig. \ref{fig:w0Tc} after continuum extrapolation.
\begin{center}
\includegraphics[width=0.75\textwidth]{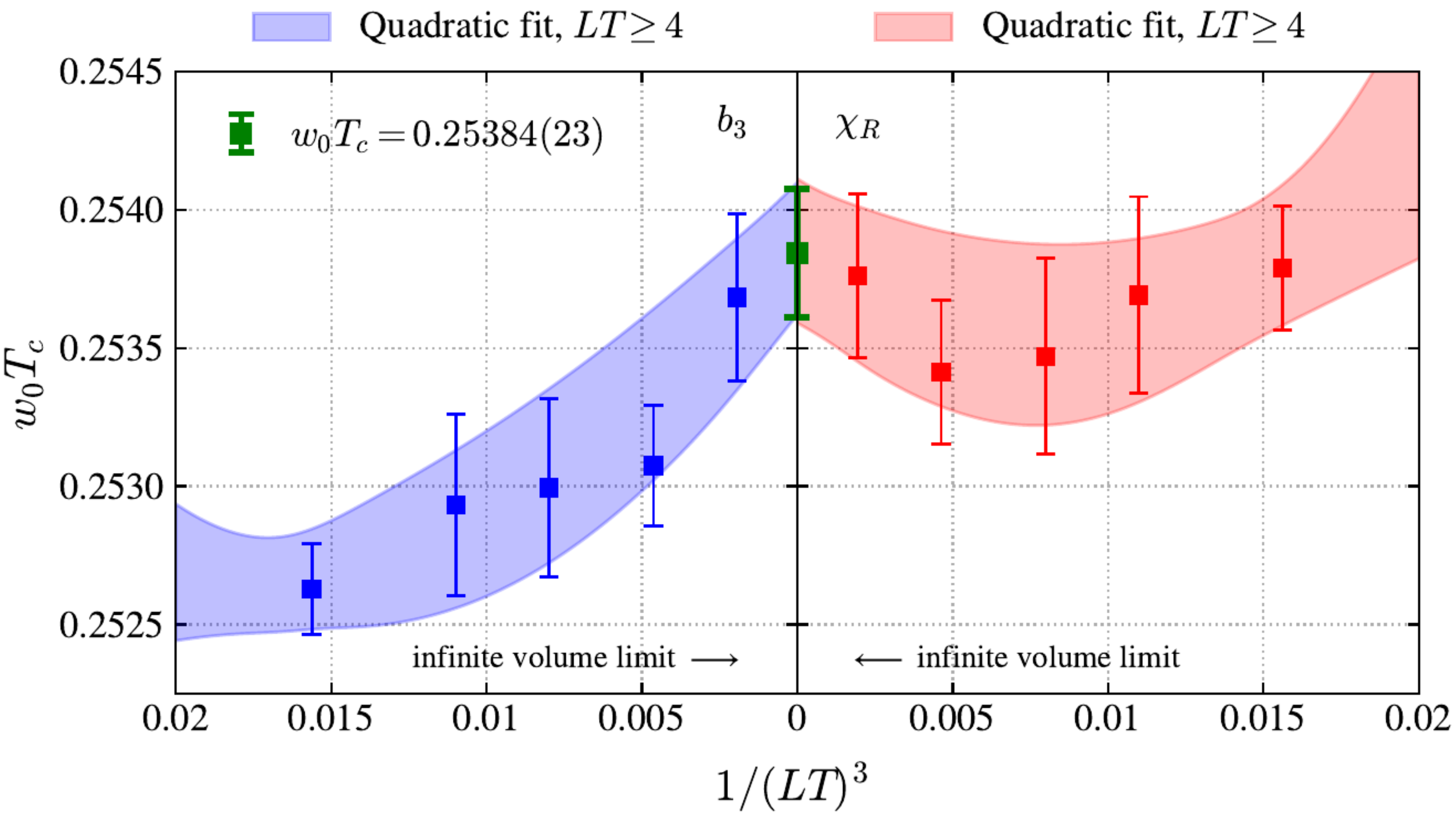}
\captionof{figure}{Continuum extrapolated transition temperature $w_0 T_c$ from $b_3=0$ (left panel) and peak of the susceptibility (right panel) as a function of the inverse physical volume $1/(LT)^3=(N_t/N_s)^3$.}
\label{fig:w0Tc}
\end{center}
\noindent
All in all 256 different analyzes \cite{prec_study} contribute to the systematic uncertainties which come from the two definitions of $\beta_c$, choice of degree of the (un-)correlated fits to the momentum with different cutoffs for the eigenvalues, WSC or SSC scale for $w_0$, degree of fit to $w_0/a$, and the range and degree of the continuum fit. We finally obtain a per-mill accurate result of 
\begin{equation}
w_0 T_c=0.25384(11)_\mathrm{stats}(21)_\mathrm{sys}.
\end{equation}
\noindent
A clear sign of a 1st order phase transition is a non-vanishing latent heat. This quantity can be seen as the gap between the energy density $\epsilon$ of both phases, which shows a discontinuity at the transition. On the lattice a suitable quantity which includes the energy density is the so-called trace anomaly defined as
\begin{equation}
\frac{\epsilon - 3p}{T^4}=N_t^4 T\frac{\partial \beta}{\partial T} \left[ S_0 - S_T \right].
\end{equation}
\noindent
It can be understood as a measure for the deviation of the system from an ideal gas. Here $S_0$ and $S_T$ are the gauge action densities at vanishing and finite temperature respectively. The trace anomaly is evaluated in the "hot" (deconfined) and "cold" (confined) phases and the difference is taken to calculate the latent heat according to
\begin{equation}
\frac{\Delta E}{T^4}:=\Delta \frac{\epsilon - 3p}{T^4}= N_t^4 T\frac{\partial \beta}{\partial T} \left[ S_\mathrm{cold} - S_\mathrm{hot} \right].
\end{equation}
\noindent
The procedure is the following: We reweight our closest ensemble to $\beta_c$ right to the critical couplings and calculate the minimum between the two peaks of the $|P|$ histogram. This minimum serves as a cut between the hot and the cold phases of the system \cite{Shirogane:2016zbf,Shirogane2021}. Then the trace anomaly is calculated for configurations whose $|P|$ is above and below the cut. One feature of this approach is the exponentially decreasing systematic error with the volume which can be seen in Fig. \ref{fig:ploop_histo_trace_anom}. Here $|P|$ histograms for various volumes are shown and one can see that the first peak, indicating the confined phase, tends to $0$ as expected. 
\begin{center}
\includegraphics[width=0.68\textwidth]{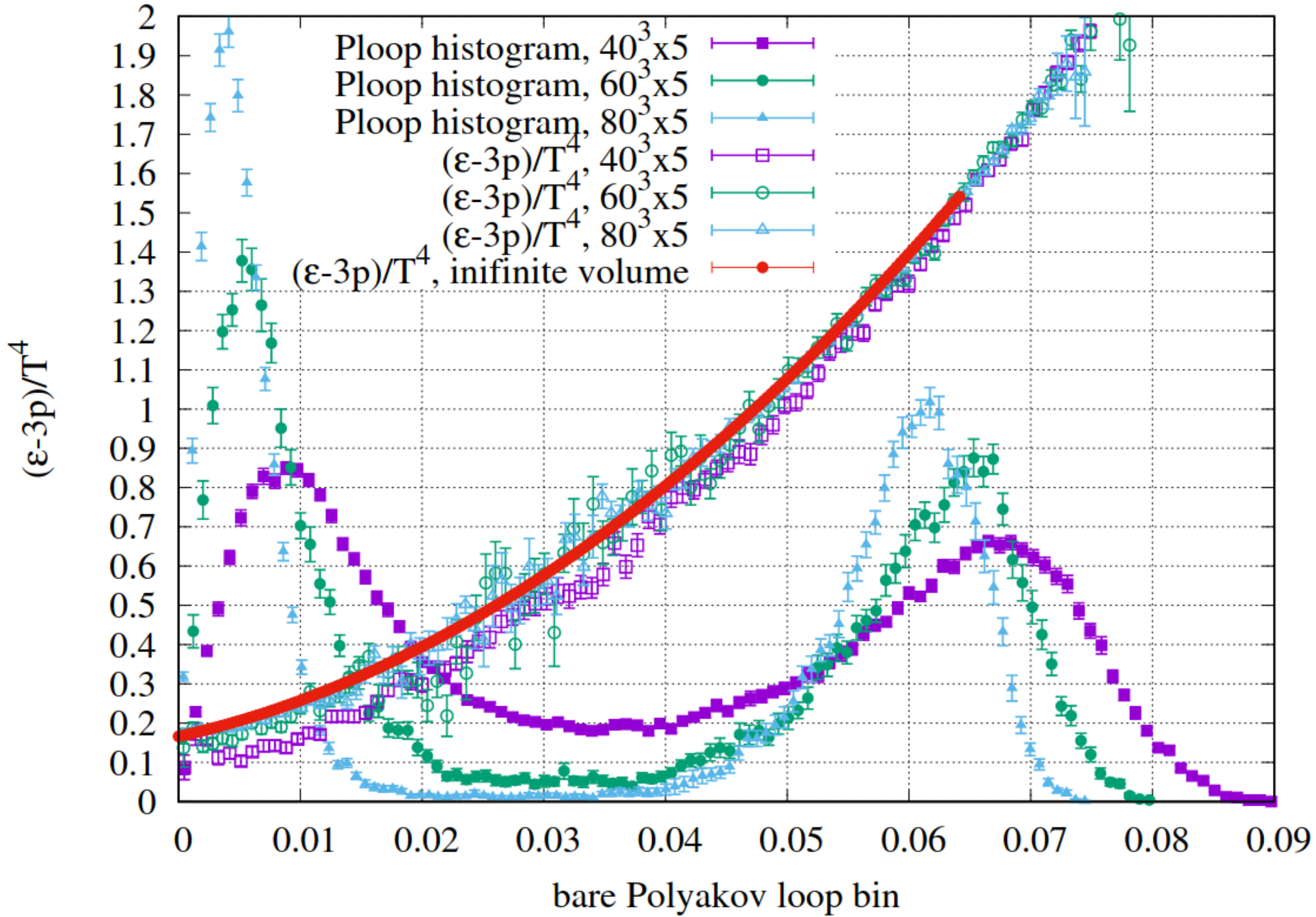}
\captionof{figure}{Histograms of the absolute value of the Polyakov loop and trace anomaly as function of the Polyakov loop bin for three different volumes.}
\label{fig:ploop_histo_trace_anom}
\end{center}
\noindent
The continuum and infinite volume extrapolations of the latent heat can be done simultaneously and the results are shown in Fig. \ref{fig:lat_heat}.
\begin{center}
\includegraphics[width=0.6\textwidth]{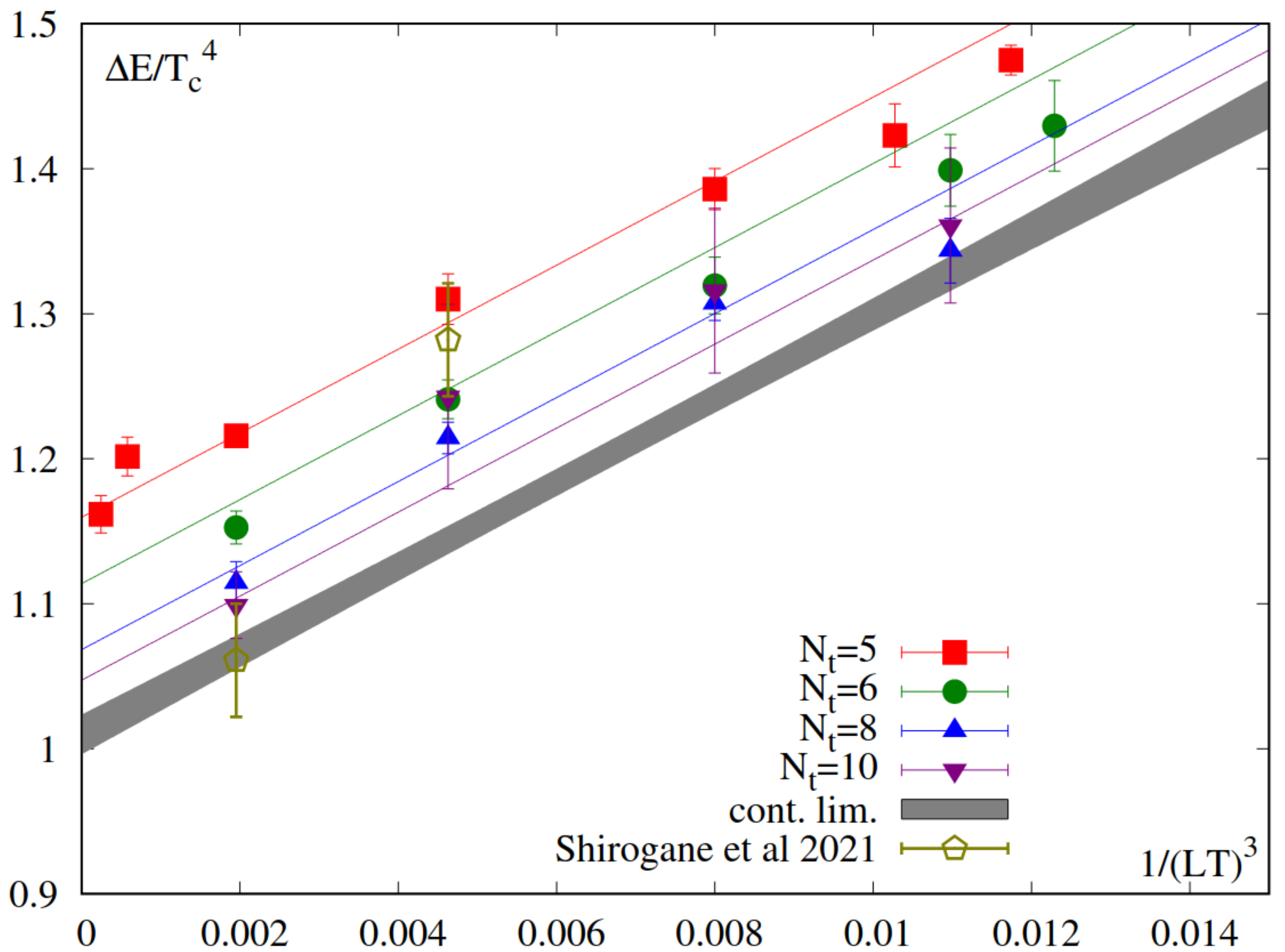}
\captionof{figure}{Latent heat as a function of the inverse physical volume for our $N_t=5,6,8,10$ ensembles and combined limit including the continuum extrapolation. In addition continuum extrapolated finite volume results from \cite{Shirogane2021} are shown.}
\label{fig:lat_heat}
\end{center}
\noindent 
Including different numbers of Polyakov loop bins, in- or excluding $N_t=5$ in the continuum extrapolation, degree of log-polynomial fit to extract the cut, in- or excluding $LT=4.5$ in the infinite volume extrapolation, using multihistogram or single-beta reweighting to extract $\beta_c$ in our systematic error analysis we get 
\begin{equation}
\Delta \left[ \frac{\epsilon - 3p}{T^4} \right] = 1.025(21)_\mathrm{stats}(27)_\mathrm{sys}.
\end{equation}
\noindent
The small but clearly non-vanishing value of the latent heat shows that thermal transition in quenched QCD is a weak 1st order transition in the context of SU(N) theories \cite{Lucini:2005vg}.
The major driver of the systematic error is the definition of $\beta_c$ and the selection of $N_t$ ensembles to the continuum limit which contribute $1.32\%$ and $1.53\%$ to the error. Further details can be found in \cite{prec_study}.

\paragraph*{Acknowledgments:}
The project received support
from the BMBF Grant No. 05P21PXFCA. The authors gratefully acknowledge the Gauss Centre for Supercomputing e.V. (www.gauss-centre.eu) for funding this
project by providing computing time on the GCS Super-
computer HAWK at HLRS, Stuttgart. Part of the computation was performed on the QPACE3 funded by the
DFG and hosted by JSC and on the cluster at the University of Graz.

\end{document}